\documentclass{icrc2009}

\usepackage{graphicx}   
\usepackage{caption}    
\usepackage[font=footnotesize]{subfig} 
\usepackage{fixltx2e}
\usepackage{url}

\newcommand{\shorttitle}[1]%
{\markboth{Proceedings of the 31\MakeLowercase{$^{st}$} ICRC, {\L}\'{o}d\'{z} 2009}{#1} }
\newcommand{\etal}{\MakeLowercase{\textit{et al. }}} 

\begin{document}
\title{The ANTARES Deep-Sea Neutrino Telescope: \\
Status and First Results}

\author{\IEEEauthorblockN{Paschal Coyle  \IEEEauthorrefmark{1} \\
On Behalf of the ANTARES Collaboration}
                            \\
\IEEEauthorblockA{\IEEEauthorrefmark{1}
coyle@cppm.in2p3.fr\\
 Centre de Physique des Particules de Marseille\\
163 Avenue de Luminy, Case 902,\\
13288 Marseille cedex 09, France}}

\shorttitle{P. Coyle \etal ANTARES status and first results}
\maketitle

\begin{abstract}

Various aspects of the construction, operation and calibration of the recently completed deep-sea ANTARES neutrino telescope are described. Some first results obtained with a partial five line configuration are presented, including depth dependence of the atmospheric muon rate, the search for point-like cosmic neutrino sources and the search for dark matter annihilation in the Sun.
 
 \end{abstract}

\begin{IEEEkeywords}
ANTARES, neutrino, point source, dark matter 
\end{IEEEkeywords}
 
\section{Introduction}
The undisputed galactic origin of cosmic rays at energies below the so-called knee implies an existence of a non-thermal population of galactic sources which effectively accelerate protons and nuclei to TeV-PeV energies. The distinct signatures of these cosmic accelerators are high energy neutrinos and gamma rays produced through hadronic interactions with ambient gas or photoproduction on intense photon fields near the source. While gamma rays can be produced also by directly accelerated electrons, high-energy neutrinos provide unambiguous and unique information on the sites of the cosmic accelerators and hadronic nature of the accelerated particles.   

ANTARES (http://antares.in2p3.fr/) is a deep-sea neutrino telescope, designed for the detection of all flavours of high-energy neutrinos emitted by both Galactic (supernova remnants, micro-quasars etc.) and extragalactic (gamma ray bursters, active galactic nuclei, etc.) astrophysical sources. The telescope is also sensitive to neutrinos produced via dark matter annihilation within massive bodies such as the Sun and the Earth. Other physics topics include measurement of neutrino oscillation parameters, the search for magnetic monopoles, nuclearites etc. 

The recently completed ANTARES detector is currently the most sensitive neutrino observatory studying the southern hemisphere and includes the particularly interesting region of the Galactic Centre in its field of view. ANTARES is also a unique deep-sea marine observatory providing continuous, high-bandwidth monitoring from a variety of sensors dedicated to acoustic, oceanographic and Earth science studies. 
 
\section{The ANTARES Detector}

  \begin{figure}[!t]
  \centering
  \includegraphics[width=2.7in]{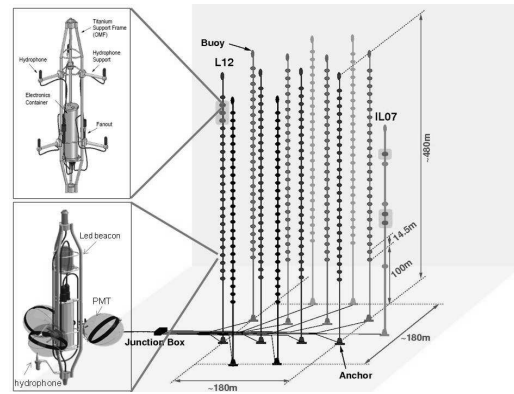}
  \caption{The layout of the completed ANTARES detector. The top insert shows an image of an acoustic storey with 
its six acoustic sensors and the lower insert an image of an optical storey with its three photomultipliers.}
  \label{fig1}
 \end{figure}

The ANTARES detector is located at a depth of 2475~m in the Mediterranean Sea, 42~km from La Seyne-sur-Mer in the South of France ($42^\circ48’ N, 6^\circ10’ E$). A schematic of the detector layout is shown in Figure \ref{fig1}. It is equipped with 885 optical sensors arranged on 12 flexible lines. Each line comprises up to 25 detection storeys each equipped with three downward-looking 10-inch photo-multipliers (PMTs), orientated at $45^\circ$ to the line axis. The lines are maintained vertical by a buoy at the top of the 450~m long line. The spacing between storeys in 14.5~m and the lines are spaced by 60-70~m. An acoustic positioning system provides real-time location of the detector elements to a precision of a few centimeters. A system of optical beacons allows in-situ time calibration. The first detection line was installed in 2006. Five lines have been operating since March 2007. Ten lines were operational in December 2007. With the installation of eleventh and twelfth lines in May 2008, the detector construction was completed. An additional line (IL07) contains an ensemble of oceanographic sensors dedicated measurement of the environmental parameters. The twelfth line and the IL07 also includes hydrophone-only storeys dedicated to the study of the ambient acoustic backgrounds; R\&D for possible acoustic detection of ultra-high energy neutrinos. 

The ANTARES Collaboration currently comprises 29 particle physics, astrophysics and sea science institutes from seven countries (France, Germany, Italy,  Netherlands, Romania, Russia and Spain). 

The three-dimensional grid of photomultiplier tubes is used to measure the arrival time and position of Cherenkov photons induced by the passage of relativistic charged particles through the sea water. The reconstruction algorithm relies on the characteristic emission angle of the light (about 43 degrees) to determine the direction of the muon and hence infer that of the incident neutrino. The accuracy of the direction information allows to distinguish upward-going muons, produced by neutrinos, from the overwhelming background from downward-going muons, produced by cosmic ray interaction in the atmosphere above the detector. Installing the detector at great depths serves to attenuate this background and also allows to operate the PMTs in a completely dark environment.   

  At high energies the large range of the muon allows the sensitive volume of the detector to be significantly greater than the instrumented volume. Although optimised for muon neutrino detection, the detector is also sensitive to the electron and tau neutrinos albeit it with reduced effective area. 

The total ANTARES sky coverage is 3.5$\pi$sr, with an instantaneous overlap of 0.5$\pi$sr with that of the Icecube experiment at the South Pole. Together ANTARES and Icecube provide complete coverage of the high-energy neutrino sky.
Compared to detectors based in ice, a water based telescope benefits from a better angular resolution, due to the absence of light scattering on dust and/or bubbles trapped in the ice. On the other hand, it suffers from additional background light produced by beta decay of  $^{40}$K salt present in the sea water as well as bioluminescent light produced by biological organisms. Furthermore, the continual movement of the detector lines, in reaction to the changing direction and intensity of the deep-sea currents, must be measured and taken into account in the track reconstruction. 

The ANTARES data acquistion \cite{bouwhuis1} is based on the 'all-data-to-shore' concept, in which all hits above a threshold of 0.3 single photon-electrons are digitised and transmitted to shore. Onshore a farm of commodity PCs apply a trigger based on requiring the presence of a 4-5 causally connected local coincidences between pairs of PMTs within a storey. The typical trigger rate is 5-10 Hz, dominated by downgoing muons. In addition, an external trigger generated by the gamma-ray bursts coordinates network (GCN) will cause all the buffered raw data (two minutes) to be stored on disk. This offers the potential to apply looser triggers offline on this subset of the data \cite{bouwhuis2}. 
 
\section{Detector Calibration}

The precision on the neutrino direction is limited at low energies by the kinematics of the neutrino interaction. For neutrino energies above 10~TeV the angular resolution is determined by the intrinisic detector resolution i.e. the timing resolution and accuracy of the location of the PMTs. The energy measurement relies on an accurate calibration of the charge detected by each PMT \cite{baret}.   

\subsection {Acoustic Positioning}
The positions of the PMTs are measured every two minutes with a high-frequency long-baseline acoustic positioning system comprising fixed acoustic emitters-receivers at the bottom of each line and acoustic receivers distributed along a line \cite{brown}. After triangulation of the positions of the moving hydrophones, the shape of each line is reconstructed by a global fit based on a model of the physical properties of the line and additional information from the tiltmeters and compass sensors located on each storey. The relative positions of the PMTs are deduced from this reconstructed line shape and the known geometry of a storey. The system provides a statistical precision of a few mm. The final precision on the PMT locations is a few cm, smaller than the physical extension of the PMTs, and  is limited by the systematic uncertainties on the knowledge of the speed of sound in the sea water.  

 \begin{figure}[!t]
  \centering
  \includegraphics[width=2.7in]{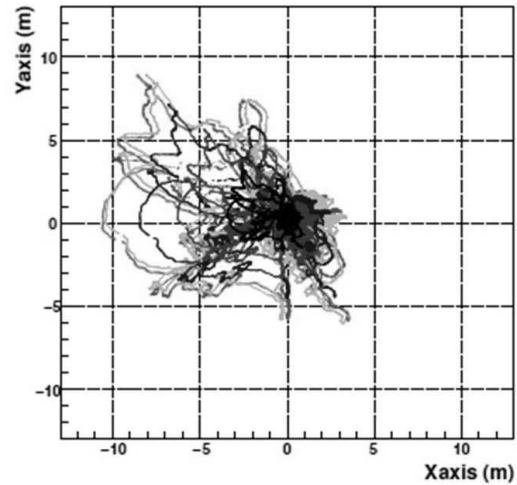}
  \caption{The horizontal movements relative to the bottom of the line, 
of all hydrophones on Line 11 for a 6 month period; black points is the
hydrophone on Storey 1, red is on Storey 8, blue is on Storey 14, green
is on Storey 22 and magenta is on Storey 25.}
  \label{fig2}
 \end{figure}

Figure \ref{fig2} shows the movements of various storeys on a line, relative to its centre axis. The extent of the displacement
depends on the intensity of the sea current. For typical currents of a few centimetres per second, the displacement is a few metres for the topmost storeys.   

\subsection {Time Calibration}
The relative time calibration of ANTARES is performed via a number of independent and redundant systems \cite{Gomez}. 
The master clock system features a method to measure the transit time of the clock signals to the electronics located in each storey of the detector. The determination of the remaining residual time offsets within a storey, due to the delays in the front-end electronics and transit time of the PMTs, are based on the detection of signals from external optical beacons distributed throughout the detector. The presence of $^{40}$K in the sea water also provides a convenient source of calibrated light which is used to verify the time offsets between the triplet of PMTs within a storey as well as study the long term stablity of the PMTs efficiencies.

Every fifth storey of a line contains an optical beacon emitting in the blue. Each beacon illuminates the neighbouring storeys on its line. Comparison of the arrival hit times within a storey provides the relative inter-storey time offsets. Intra-storey time offsets can also be established after corrections are applied for time walk and 'first photon' effects.

\begin{figure}[!t]
  \centering
  \includegraphics[width=2.7in]{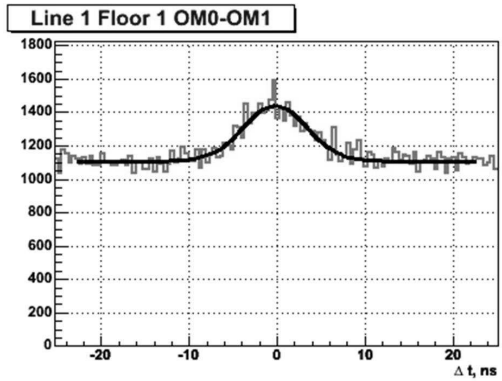}
  \caption{Conicidence peak due to  $^{40}$K decays for a single pair of photomultipliers in one storey.}
  \label{fig3}
 \end{figure}

Potassium-40 is a radioactive isotope naturally present in the sea water. The decay $^{40}K \rightarrow  e^- \nu_e ^{40} Ca$ yields an electron with an energy up to $1.3$~MeV. This energy exceeds the Cherenkov threshold for electrons in water 
(0.26~MeV), and is sufficient to produce up to 150 Cherenkov photons. If the decay occurs in the vincinity of a detector storey, a coincident signal may be recorded by pairs of PMTs on the storey. In Figure \ref{fig3} the distribution of the measured time difference between hits in neighbouring PMTs of the same storey is shown.  The peak around 0~ns is mainly due to single $^{40}$K decays producing coincident signals. The fit of the data is the sum of a Gaussian distribution and a flat background. The full width at half maximum of the Gaussian function is about 9~ns. This width is mainly due to the spatial distribution of the  $^{40}$K decays around the storey. The positions of the peaks of the time distributions for different
pairs of PMTs in the same storey are used to cross-check the time offsets computed by the optical beacon system. This is illustrated in Figure  \ref{fig3} which shows a comparision of the time offsets calculated by the optical beacons and that extracted from the  $^{40}$K analysis; an rms of 0.6~ns is obtained.  

\begin{figure}[!t]
  \centering
  \includegraphics[width=2.7in]{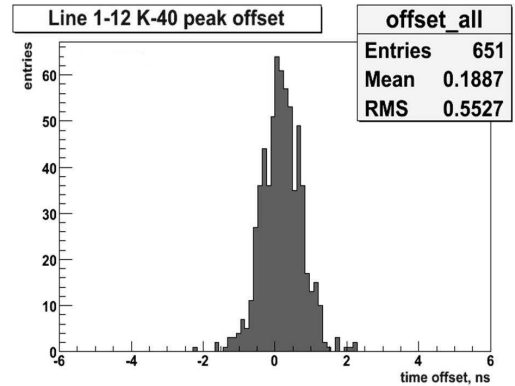}
  \caption{Time offsets for all photomultipliers as extracted using the LED beacons and independently
checked by the K40 conicidence method.}
  \label{fig4}
 \end{figure}

The rate of genuine  $^{40}$K coincidences is given by the integral under the peak of Figure \ref{fig3} and is used to monitor the relative efficiencies of all PMTs and their temporal stability.

\section{Muon Reconstruction}

\begin{figure}[!t]
  \centering
  \includegraphics[width=2.7in]{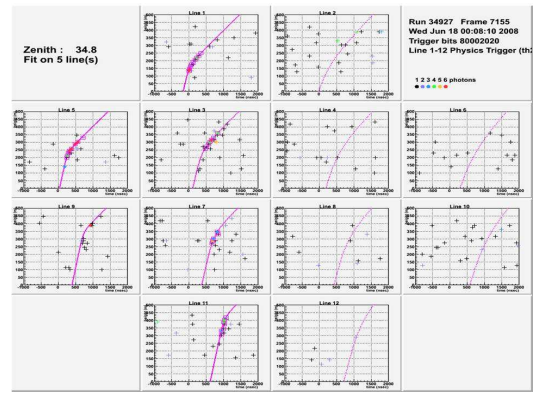}
  \caption{Example display of a neutrino candidate. The 2D plots, one for each line, show on the y axis the vertical position of the PMT with a hit and on the x axis the arrival time of the hit. The fit to the arrival time distribution corresponds to the chi-square algorithm.}
  \label{fig5}
 \end{figure}

Two alternative algorithms for reconstruction of the muon trajectories have been developed \cite{heijboer}. 
In the first approach, a simple chi-square fit is applied to a high purity sample of pre-selected hits. In addition, this algorithm merges the hits observed by the PMTs of the triplet and assumes that they are located on the line axis, i.e. the azimuthal orientation of the storey, measured by the compasses, is ignored. In  Figure \ref{fig5} an upgoing neutrino candidate fitted using this algorithm is shown. This algorithm was initially adopted as a fast reconstruction for online monitoring of the detector. Although it provides an non-optimum angular resolution (typically 1-2 degrees above 10~TeV) it has been used in a number of analyses for which the ultimate angular resolution is not crucial.

In the second approach a full maximum likelihood fit is applied, which uses a detailed parameterisation, derived from simulation, of the probability density function for the track residuals. The fit includes most hits in the event and the PDF takes into account the probability that photons arrive late due to Cherenkov emission by secondary particles or light scattering. A number of increasingly sophisticated prefits are used to aid in the location of the correct maxima of the likelihood. This algorithm makes use of the maximum amount of information, including the line shape and storey orientation, and provides
an angular resolution better than 0.3 degrees above 10~TeV.  
 
\section{Atmospheric Muons}

The dominant signal observed by ANTARES is due to the passage of downgoing atmospheric muons, whose 
flux exceeds that of neutrino-induced muons by several orders of magnitude. They are produced by high
energy cosmic rays interacting with atomic nuclei of the upper atmosphere, producing charged pions and
kaons, which subsequently decay into muons. Although an important background for neutrino detection, they are 
useful to verify the detector response. In particular, with three years of data taking, a deficit in the muon flux in the direction of the moon should allow an important verification of the pointing accuracy of the detector. 

\begin{figure}[!t]
  \centering
  \includegraphics[width=2.7in]{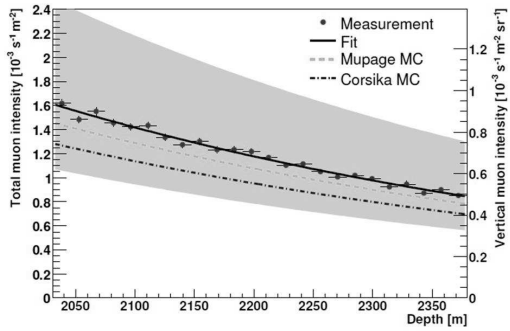}
  \caption{Attenuation of the flux of muons as a function of depth, as extracted using the adjacent storey coincidence method. The shaded band represents the systematic uncertainties due to detector effects. Predictions from MUPAGE and Corsika Monte Carlo simulations are also shown.}
  \label{fig6}
 \end{figure}

\begin{figure}[!t]
  \centering
  \includegraphics[width=3.1in]{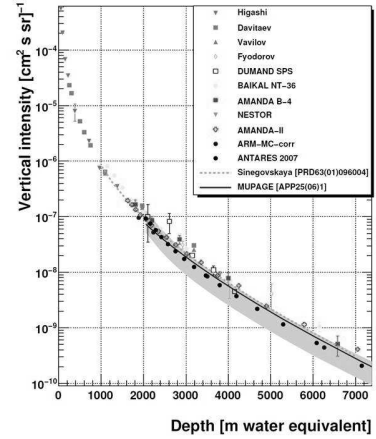}
  \caption{Vertical depth intensity relation of atmospheric muons with $E_\mu > 20$~GeV (black points). The
error band represents systematic uncertainties. A compliation of data from other experiments are also shown.}
  \label{fig7}
 \end{figure}

Two different studies of the vertical depth intensity relation of the muon flux have been performed. In the first, the attenuation of the muon flux as a function of depth is observed as a reduction in the rate of coincidences between adjacent storeys along the length of the detection lines \cite{zaborov}. This method has the advantage that it does not rely on any track reconstruction. In Figure \ref{fig6} the depth dependence of the extracted flux for the 24 inter-storey measurements averaged over all lines is shown.

In the second study, a full track reconstruction is performed and the reconstructed zenith angle is converted to an equivalent slant depth through the sea water \cite{bazotti}. Taking into account the known angular distribution of the incident muons, a depth intensity relation can be extracted (Figure \ref{fig7}). The results ae in reasonable agreement with previous measurements.

\begin{figure}[!t]
  \centering
  \includegraphics[width=2.7in]{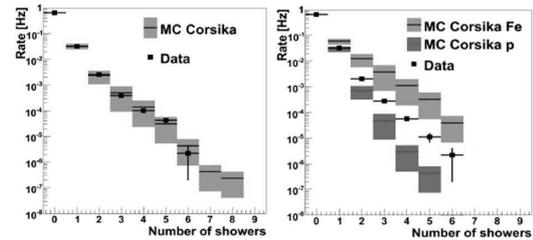}
  \caption{Identification of energetic electromagnetic showers: (left) Data and Monte Carlo comparison of the number of reconstructed electromagnetic showers for the 5-line data, assuming the nominal primary cosmic-ray composition. (right) dependence on primary cosmic-ray composition (proton versus iron).}
  \label{fig8}
 \end{figure}

The composition of the primary cosmic rays in the knee region is of particular interest. As the number of identified electromagnetic showers in an event depends on the muon energy and the number of muons present in a muon bundle, it is sensitive to the primary cosmic ray composition. An algorithm has been developed to  estimate the number of energetic electromagnetic (EM) showers generated along the muon trajectory \cite{mangano1}. This algorithm relies on the fact that the emission point of Cherenkov photons from the muon are uniformly distributed along the muon trajectory whereas Cherenkov photons orginating from an electromagnetic shower will tend to cluster from a single point. The efficiency and purity of the algorithm to identify a shower depends on the shower energy, for example the efficiency to identify a 1~TeV shower is 20\% with a purity of 85\%. In Figure \ref{fig8} (left) the distribution of the number of reconstructed energetic showers per event in the 5-line data is shown. Good agreement with the Corsika Monte Carlo is obtained when the 22Horandel primary composition model is assumed. In Figure \ref{fig8} (right) the data distribution is compared with that obtained assuming a pure proton or a pure iron primary cosmic ray composition. 

A search for a large-scale anistropy in the arrival directions of the atmospheric muons  has been performed but with the current statistics is not yet sensitive to the $0.1\%$ level variations reported by other experiments \cite{mangano2}. The possibilty for detection of gamma ray induced air showers with ANTARES is also under evaluation \cite{guillard}. 

\section{Search for Cosmic Neutrino Point Sources}

\begin{figure}[!t]
  \centering
  \includegraphics[width=2.7in]{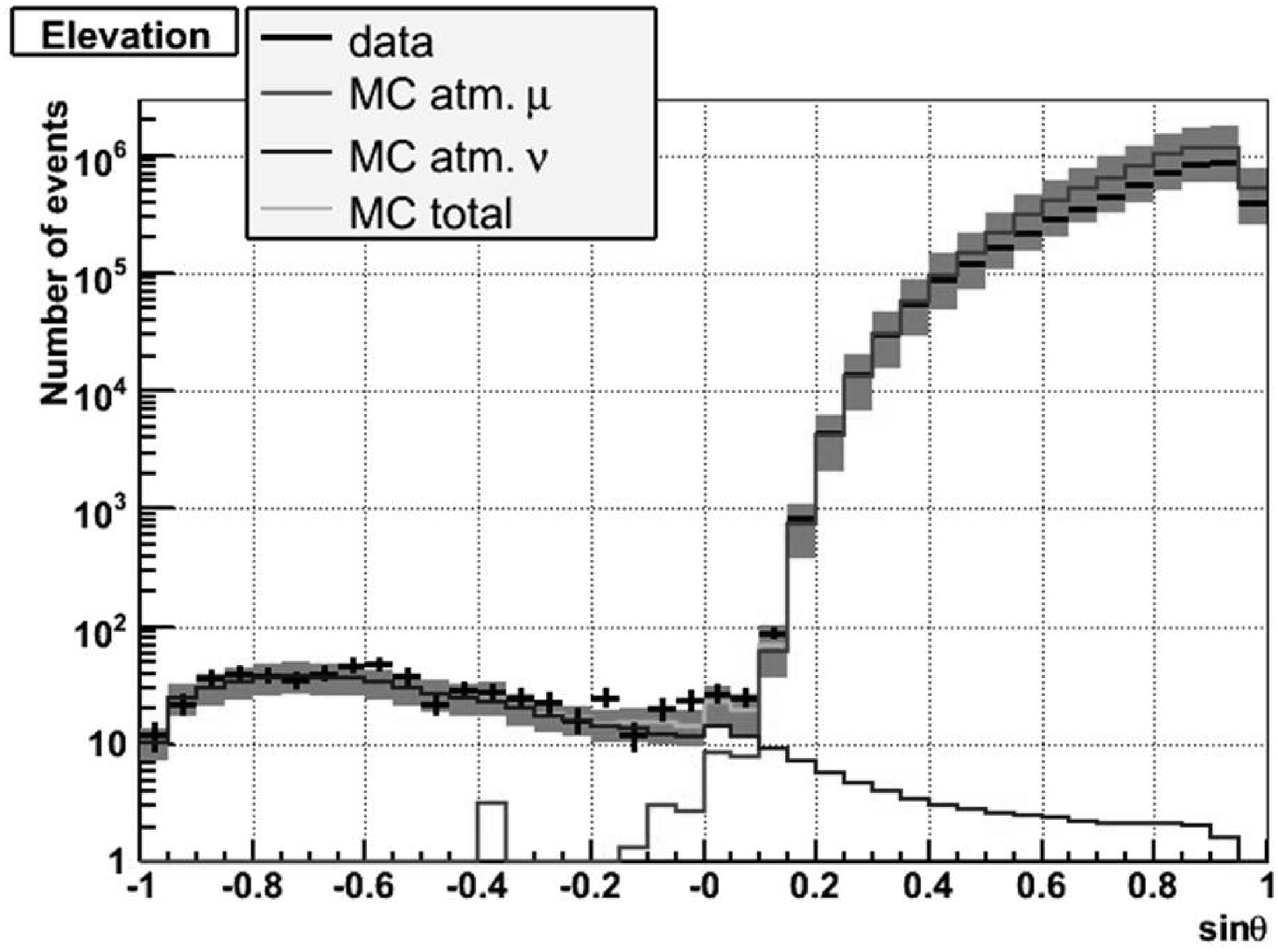}
  \caption{Zenith distribution of reconstructed muons in the 2008 data. The Mone Carlo expectation for the atmospheric muon and atmospheric neutrino backgrounds are indicated.}
  \label{fig9}
 \end{figure}

The muons produced by the interaction of neutrinos can be isolated from the muons generated by the cosmic ray interactions by requiring that the muon trajectory is reconstructed as up-going. In Figure \ref{fig9} the zenith angular distribution of muons in the 2007+2008 data sample by the $\chi^2$ reconstruction algorithm is shown. A total of 750 mulitline up-going neutrinos candidates are found, in good agreement with expectations from the atmospheric neutrino background.

\begin{figure}[!t]
  \centering
  \includegraphics[width=2.7in]{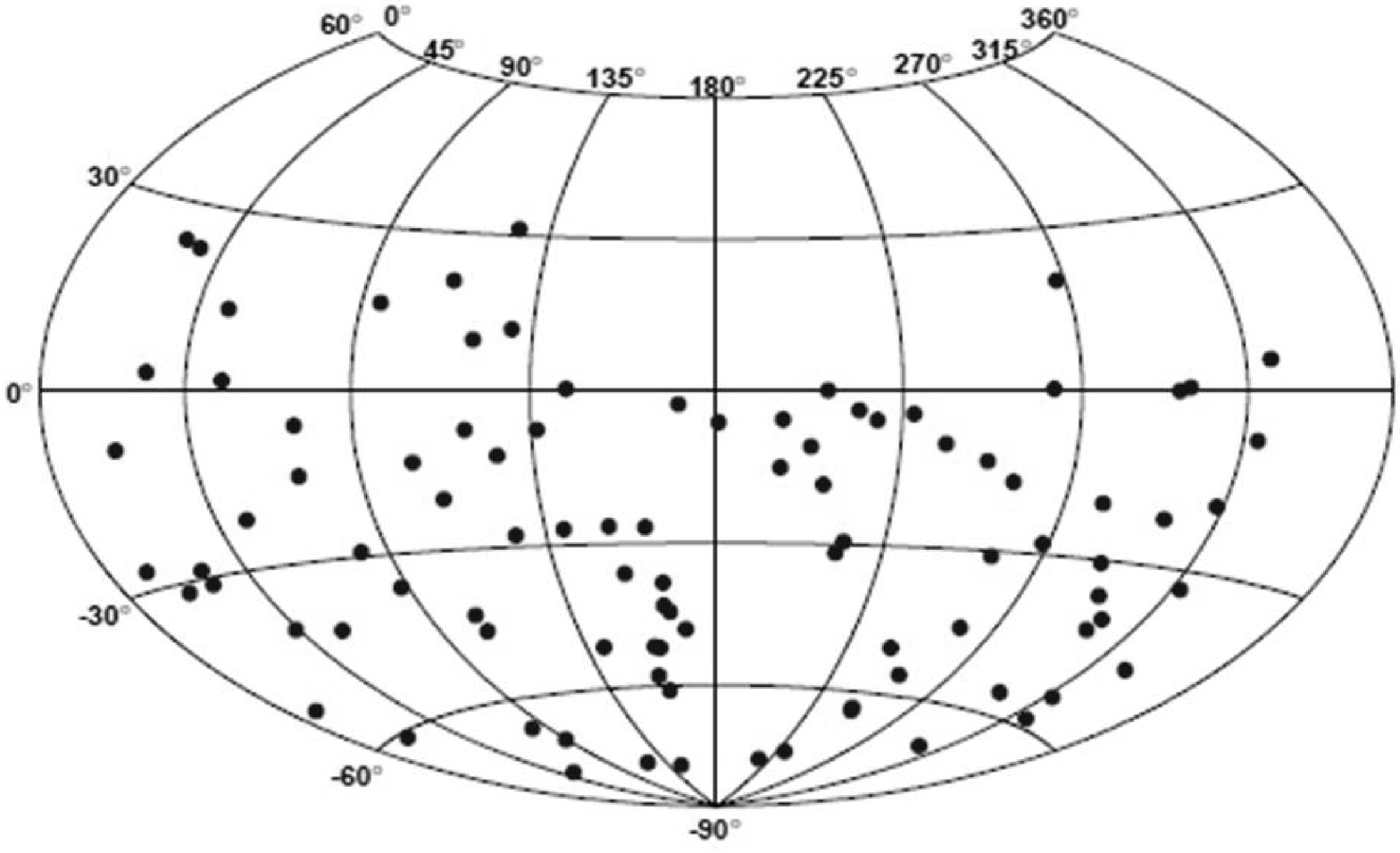}
  \caption{Sky map, in geocentric coordinates, of the upgoing neutrino candidates for the 2007 data.}
  \label{fig10}
 \end{figure}

\begin{table}
 \begin{center}
{\footnotesize

\begin{tabular}{|l|r@{.}l|r@{.}l|c|c|c|} \hline 
Source name &    \multicolumn{2}{c|}{$\delta$ ($^{\circ}$)}  &  \multicolumn{2}{c|}{RA  ($^{\circ}$)} & n$_{{\rm bin.}}$ &  p-value &    $\phi_{90}$ \\ \hline \hline
PSR B1259-63      &    -63&83 &    195&70  & 0 &     -  &    3.1  \\
RCW 86            &    -62&48 &    220&68  & 0 &     -  &    3.3  \\
HESS J1023-575    &    -57&76 &    155&83  & 1 &  0.004 &    7.6  \\
CIR X-1           &    -57&17 &    230&17  & 0 &     -  &    3.3  \\
HESS J1614-518    &    -51&82 &    243&58  & 1 &  0.088 &    5.6  \\
GX 339            &    -48&79 &    255&70  & 0 &     -  &    3.8  \\
RX J0852.0-4622   &    -46&37 &    133&00  & 0 &     -  &    4.0  \\
RX J1713.7-3946   &    -39&75 &    258&25  & 0 &     -  &    4.3  \\
Galactic Centre   &    -29&01 &    266&42  & 1 &  0.055 &    6.8  \\
W28               &    -23&34 &    270&43  & 0 &     -  &    4.8  \\
LS 5039           &    -14&83 &    276&56  & 0 &     -  &    5.0  \\
HESS J1837-069    &    -6&95  &    279&41  & 0 &     -  &    5.9  \\
SS 433            &     4&98  &    287&96  & 0 &     -  &    7.3  \\
HESS J0632+057    &     5&81  &    98&24   & 0 &     -  &    7.4  \\ \hline
ESO 139-G12       &    -59&94 &    264&41  & 0 &     -  &    3.4  \\
PKS 2005-489      &    -48&82 &    302&37  & 0 &     -  &    3.7  \\
Centaurus A       &    -43&02 &    201&36  & 0 &     -  &    3.9  \\
PKS 0548-322      &    -32&27 &    87&67   & 0 &     -  &    4.3  \\
H 2356-309        &    -30&63 &    359&78  & 0 &     -  &    4.2  \\ 
PKS 2155-304      &    -30&22 &    329&72  & 0 &     -  &    4.2  \\
1ES 1101-232      &    -23&49 &    165&91  & 0 &     -  &    4.6  \\
1ES 0347-121      &    -11&99 &    57&35   & 0 &     -  &    5.0  \\
3C 279            &    -5&79  &    194&05  & 1 &   0.030&    9.2  \\
RGB J0152+017     &     1&79  &    28&17   & 0 &     -  &    7.0  \\ \hline
IC22 hotspot      &    11&4  &    153&4  & 0 &     -  &    9.1  \\ \hline

\end{tabular}
}
 \caption{\small Results of the search for cosmic neutrinos correlated
 with potential neutrino sources. The sources are divided into three
 groups: galactic (top), extra-galactic (middle) and the hotspot from
 IceCube with 22 lines (bottom). The source name and location in
 equatorial coordinates are shown together with the number of events
 within the optimum cone for the binned search, the p-value of the
 unbinned method (when different from 1) and the corresponding upper
 limit at 90\% C.L.  $\phi_{90}$ is the value of the normalization
 constant of the differential muon-neutrino flux assuming an $E^{-2}$
 spectrum (i.e. $E^{2} d\phi_{\nu_{\mu}} / dE \le \phi_{90} \times
 10^{-10} $ TeV cm$^{-2}$ s$^{-1}$). The integration energy range is 10
 GeV - 1 PeV.}
 \label{tab:sources}
 \end{center}
\end{table}

For a subset of this data (the 5-line data of 2007) the angular resolution has been improved by applying the pdf based track reconstruction, which makes full use of the final detector alignment. After reoptimisation of the selection cuts for the best upper limits, 94 upgoing neutrino candidates are selected \cite{toscano}. The corresponding sky map for these events is shown in Figure \ref{fig10}. An all sky search, independent of assumption on the source location, has been performed on these data. The most significant cluster was found at ($\delta=-63.7^\circ, RA=243.9^\circ$) with a corresponding p-value of 0.3 ($1\sigma$ excess) and is therefore not significant. 

\begin{figure}[!t]
  \centering
  \includegraphics[width=3.1in]{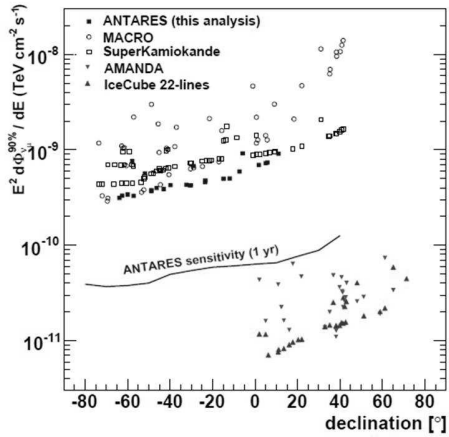}
  \caption{Neutrino flux upper limits at $90\%$ C.L. obtained by this analysis (solid squares), compared with published results from other experiments (IceCube [24], AMANDA [25], SuperKamiokande [26] and MACRO [27]). The expected sensitivity of ANTARES for one year with twelve lines is also shown (solid line). The source spectrum assumed in these results is $E^{−2}$, except for MACRO, for which an $E^{−2.1}$ spectrum was used.}
  \label{fig11}
 \end{figure}

A search amongst a pre-defined list of 24 of the most promising galactic and extra-galactic neutrino sources (supernova remnants, BL Lac objects, Icecube hot spot, etc.) is reported in Table \ref{tab:sources}. The lowest p-value obtained (a $2.8 \sigma$ excess, pre-trial) corresponds to the location ($\delta=-57.76^\circ, RA=155.8^\circ$). Such a probablity or higher would be expected in $10\%$ of background-only experiments and is therefore not significant. The corresponding flux upper limits, assuming an $E^{-2}$ flux, are plotted in Figure \ref{fig11} and compared to published upper limits from other experiments. Also shown in Figure \ref{fig11} is the predicted upper limit for ANTARES after one full year of twelve line operation.

\section{Search for Dark Matter}
In many theoretical models a Weakly Interacting Massive Particle (WIMP), a relic from the Big Bang, is proposed to explain the formation of structure in the universe and the discrepancy observed between the measured rotation curves of stars and the associated visible matter distribution in galaxies. A generic property of such WIMPs is that they gravitationally accumulate at the centre of massive bodies such as the Sun or the Earth, where they can self annihilate into normal matter. Only neutrinos, resulting from the decay of this matter, can escape from the body and be detected by a neutrino telescope.

\begin{figure}[!t]
  \centering
  \includegraphics[width=2.7in]{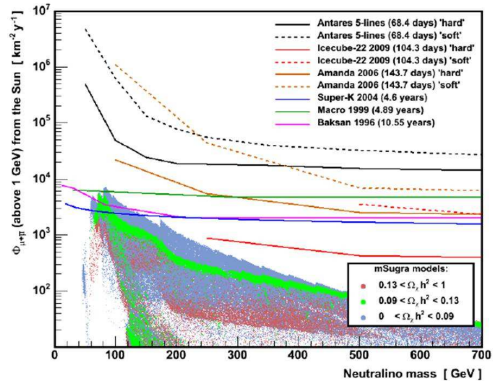}
  \caption{Upper limit on the muon flux from the Sun as a function of neutralino mass. The expected fluxes for a scan of mSUGRA parameters is shown as well as a variety of limits from other experiments.}
  \label{fig12}
 \end{figure}

\begin{figure}[!t]
  \centering
  \includegraphics[width=2.7in]{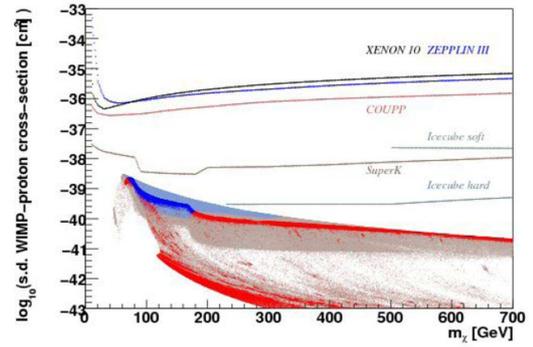}
  \caption{The spin dependent WIMP-proton cross-section versus neutralino mass in the mSUGRA model. The points 
are the results of the scan over the range of model parameters. The points in blue would be excluded at 90\% CL after three years of ANTARES operation. Existing upper limits from a variety of direct and indirect direction experiments are shown. }
  \label{fig13}
 \end{figure}

Within Supersymmetric models with R-parity conservation, the lightest supersymmetric particle (LSP) is stable and is the WIMP candidate. In order to predict the expected solar neutrino fluxes the constrained phenomenological framework of the minimal Supergravity model (mSUGRA, computations using ISASUGRA[5]) has been adopted. Figure \ref{fig10} shows the predicted integrated neutrino fluxes above 10~GeV in ANTARES as a function of neutralino mass for the scan of the model parameters: scalar mass $m_0$ in [0,8000] GeV, gaugino mass $m_{1/2}$ in [0,2000] GeV, tri-linear scalar coupling $A_0$ in [$-3m_0,3m_0$], sign of the Higgsino mixing parameter: $\mu > 0$, ratio of Higgs fields vacuum expectation values $tan \beta$ in [0,60], $m_{top}=172.5~GeV$. The local Dark Matter halo density (NFW-model) was set to $0.3~GeV/cm^3$. 
The most favourable models for neutrino telescopes are in the so-called ‘focus point’ region, for which the decays are mainly via $W^+W^-$ leading to a harder neutrino spectra. Thanks to its low-energy threshold ANTARES is ideally suited to address  low-mass neutralino scenarios.   

A search for neutrinos from the direction of the Sun in the 5-line data \cite{lim}, showed no excess with respect to background expectations. The corresponding derived limit on the neutrino flux is shown in Figure \ref{fig12}. Also shown is the expected limit with 5 years of data taking with the full 12-line detector; a large fraction of the focus point region could be excluded.
  
Due to the capture of the WIMPs inside the Sun, which is mainly hydrogen, neutrino telescopes are particularly sensitive to the spin-dependent coupling of the WIMPs to standard matter. In Figure \ref{fig13} the corresponding limit on the spin-dependent WIMP-proton cross-section after three years of ANTARES operation is shown. For this case, the neutrino telescopes are significantly more sensitive than the direct direction experiments. 

\section{Multi-Messenger Astronomy}

\begin{figure}[!t]
  \centering
  \includegraphics[width=2.7in]{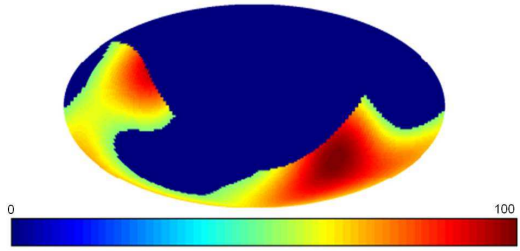}
  \caption{Common sky coverage for VIRGO/LIGO/ANTARES in geocentric coordinates. This map shows the combined antenna pattern for the gravitational wave detector network (above half-minimum), assuming that ANTARES has 100\% visibility in its antipodal hemisphere and 0\% elsewhere.}
  \label{fig14}
 \end{figure}

In order to augment the discovery potential of ANTARES, a program of collaboration with other types of observatory have been established. In this ''multi-messenger'' approach the detection threshold can be lowered in each separate experiment while preserving an acceptable rate of accidental coincidences. One example of such a program is being pursued with the gravitational wave detectors VIRGO and LIGO \cite{elewyck}. Both of these detectors had a data-taking phase during 2007 (VIRGO science run 1 and LIGO S5) which partially coincided with the ANTARES 5-line configuration. A new common science run has also recently started in coincidence with the ANTARES 12-line operation. The common sky coverage for ANTARES-VIRGO+LIGO is signifcant and is shown in Figure \ref{fig14}.
  
In a similar vein, a collaboration with the TAROT optical telescopes has been established \cite{dornic}. The directions of interesting neutrino triggers (two neutrinos within 3 degrees within a time window of 15 minutes or a single neutrino of very high energy) are sent to the Chile telescope in order that a series of optical follow up images can be taken. This procedure is well suited to maximise the sensitivity for transient events such as gamma ray bursters or flaring sources. 

\section {Acoustic Detection}

\begin{figure}[!t]
  \centering
  \includegraphics[width=2.7in]{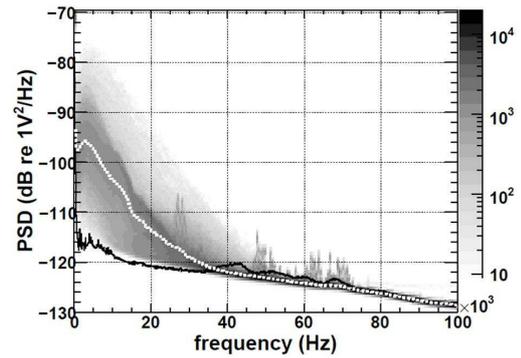}
  \caption{Power spectral density (PSD) of the ambient noise recorded with one sensor. Shown in grey shades of grey is the occurence rate in arbirary units, where dark colours indicate higher occurence. Shown as a white dotted line is the mean value of the in-situ PSD and as a black solid line the noise level recorded in the laboratory before deployment.}
  \label{fig15}
 \end{figure}

\begin{figure}[!t]
  \centering
  \includegraphics[width=2.7in]{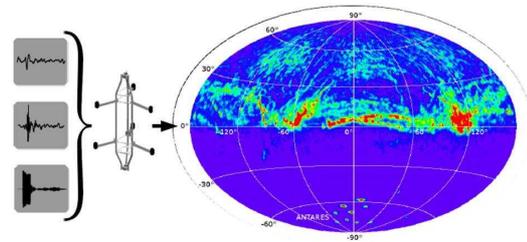}
  \caption{Map of the angular directions of the detected transient acoustic signals at the ANTARES site. }
  \label{fig16}
 \end{figure}

Due to the large attenuation length, $\approx$ 5~km for 10~kHz signals, the detection of bipolar acoustic pressure pulses  in huge underwater acoustic arrays is a possible approach for the identification of cosmic neutrinos with energies exceeding 100~PeV. To this end, the ANTARES infrastructure incorporates the AMADEUS system, an R\&D project intended to evaluate the acoustical backgrounds in the deep sea \cite{simeone}. It comprises a set of piezo-electric sensors for broad-band recording of acoustic signals with frequencies ranging up to 125~kHz with typical sensitivities around 1V/$\mu$Pa. The sensors are distributed in six ''acoustic clusters'', each comprising six acoustic sensors that are arranged at distances roughly 1~m from each other. The clusters are installed along the line 1é and the IL07 line at a horizontal distance of 240~m. The vertical spacing within a line range from 15~m to 125~m (see Figure \ref{fig1}). 

In Figure \ref{fig15} is shown the measured power spectral density of the ambient noise recorded with an acoustic sensor. Due to wind generated surface noise, the observed noise level is larger than that measured in the laboratory. Strong correlation of the measured acoustic noise with the measured surface wind speed are observed. 

In Figure \ref{fig16} an acoustic "sea map" of all transient (not just bipolar) signals detected by the apparatus during a one month period is shown. The acoustic pinger of the ANTARES positioning system are clearly identified in the lower hemisphere. The acoustic activity in the upper hemisphere is presumably due to surface boats and possibly marine mammals.  

\section{Conclusion}

After more than a decade of R\&D and prototyping the construction of ANTARES, the first operating deep-sea neutrino telescope has been completed. Since the deployment of the first line in 2006, data taking has proceeded essentially continuously. During this time the methods and procedures to calibrate such a novel detector have been developed, including in-situ time calibration with optical beacons and acoustic positioning of the detector elements with acoustic hydrophones. The presence of the $^{40}K$ in the sea water has proven particularly useful for monitoring the stability of the time calibration as well as the detector efficiency.

Based on data from the intermediate 5-line configuration, a number of preliminary analyses have been presented;  measurements of the atmospheric muon vertical depth intensity relation, a search for cosmic neutrino point sources in the southern sky, and a search for dark matter annihilation in the Sun. For the latter two analyses no significant signal was observed and competitive upper limits have been obtained. 

The succesful operation of ANTARES, and analysis of its data, is an important step towards KM3NET \cite{km3net}, a future km$^3$-scale high-energy neutrino observatory and marine sciences infrastructure planned for construction in the Mediterranean Sea.


\newpage
 
 \begin{thebibliography}{99}

   \bibitem{bouwhuis1}   M.~Bouwhuis, \emph{Concepts and performance of the Antares data acquisition system}, ICRC2009, arXiv:0908.0811.

  \bibitem{bouwhuis2}   M.~Bouwhuis,  \emph{Search for gamma-ray bursts with the Antares neutrino telescope}, ICRC2009, 
arXiv:0908.0818.

  \bibitem{baret}   B.~Baret,  \emph{Charge Calibration of the ANTARES high energy neutrino telescope}, ICRC2009, arXiv:0908.0810.

  \bibitem{brown}   A.~Brown, \emph{Positioning system of the ANTARES Neutrino Telescope}, ICRC2009, arXiv:0908.0814.

  \bibitem{Gomez}   P.~Gomez, \emph{Timing Calibration of the ANTARES Neutrino Telescope}, ICRC2009, arXiv:0911.3052.

  \bibitem{heijboer}   A.~Heijboer,  \emph{Reconstruction of Atmospheric Neutrinos in Antares}, ICRC2009, arXiv:0908.0816.

  \bibitem{zaborov}   D.~Zaborov, \emph{Coincidence analysis in ANTARES: Potassium-40 and muons}, 
43rd Rencontres de Moriond on Electroweak Interactions and Unified Theories, La Thuile, Italy, 1-8 Mar 2008, arXiv:0812.4886

  \bibitem{bazotti}   M.~Bazzotti, \emph{Measurement of the atmospheric muon flux
with the ANTARES detector}, ICRC2009, arXiv:0911.3055.

  \bibitem{mangano1}   M.~Mangano, \emph{Algorithm for muon electromagnetic shower reconstruction}, Nucl. Instrum. Meth. A588:107-110, 2008, arXiv:0711.3158.

 \bibitem{mangano2}   M.~Mangano,  \emph{Skymap for atmospheric muons at TeV energies measured in deep-sea neutrino telescope ANTARES}, ICRC2009, arXiv:0908.0858.

 \bibitem{guillard}   G.~Guillard,  \emph{Gamma ray astronomy with Antares}, ICRC2009, arXiv:0908.0855.

 \bibitem{toscano}   S.~Toscano, \emph{Point source searches with the ANTARES neutrino telescope}, ICRC2009, arXiv:0908.0864.

 \bibitem{lim}   G.~Lim,\emph{ First results on the search for dark matter in the Sun with the ANTARES neutrino telescope}, ICRC2009, arXiv:0905.2316.
 
 \bibitem{elewyck}   V.~Elewyck, \emph{Searching for high-energy neutrinos in coincidence with gravitational waves with the ANTARES and VIRGO/LIGO detectors}, ICRC2009, arXiv:0908.2454.

 \bibitem{dornic}   D.~Dornic, \emph{Search for neutrinos from transient sources with the ANTARES telescope and optical follow-up observations}, ICRC2009, arXiv:0908.0804.

 \bibitem{simeone}   M.~Simeone, \emph{Underwater acoustic detection of UHE neutrinos with the ANTARES experiment}, ICRC2009, arXiv:0908.0862.

\bibitem{km3net} J. P. Ernenwein, \emph{Status of the KM3NeT project}, this conference, ICRC2009.

\end{thebibliography}
\end{document}